# Automated Data Integration, Cleaning and Analysis Using Data Mining and SPSS Tool For Technical School in Malaysia

**Tajul Rosli Razak** *tajulrosli@perlis.uitm.edu.my*
*Faculty of Computer and Mathematical Sciences*
*Universiti Teknologi MARA(Perlis)*
*02600 Arau, Perlis, Malaysia.*

**Abdul Hapes Mohammed** *hapes232@perlis.uitm.edu.my*
*Faculty of Computer and Mathematical Sciences*
*Universiti Teknologi MARA(Perlis)*
*02600 Arau, Perlis, Malaysia.*

**Noorfaizalfarid Hj Mohd Noor** *noorfaizal455@perlis.uitm.edu.my*
*Faculty of Computer and Mathematical Sciences*
*Universiti Teknologi MARA(Perlis)*
*02600 Arau, Perlis, Malaysia.*

**Muhamad Arif Hashim** *muhamadarif487@perlis.uitm.edu.my*
*Faculty of Computer and Mathematical Sciences*
*Universiti Teknologi MARA(Perlis)*
*02600 Arau, Perlis, Malaysia.*

**Abstract**

Students' performance plays major role in determining the quality of our education system. Sijil Pelajaran Malaysia (SPM) is a public examination compulsory to be taken by Form 5 students in Malaysia. The performance gap is not only a school and classroom issue but also a national issue that must be addressed properly. This study aims to integrate, clean and analysis through automated data mining techniques. Using data mining techniques is one of the processes of transferring raw data from current educational system to meaningful information that can be used to help the school community to make a right decision to achieve much better results. This proved DM provides means to assist both educators and students, and improve the quality of education. The result and findings in the study show that automated system will give the same result compare with manual system of integration and analysis and also could be used by the management to make faster and more efficient decision in order to map or plan efficient teaching approach for students in the future.

## 1. INTRODUCTION

Examinations serve many purposes, which are to make assessment on the effectiveness of our education process, and subsequently facilitate improvement on the process. Examinations also serve the function of differentiations among students so that different groups of student with unique level learning ability can be grouped together for differentiated education.

Education is viewed as a critical factor in contributing to the long-term economic well-being of the country. Therefore, government realizes that the importance of maximizing the potential of each individual student, as well as the education system. In Malaysia, students generally are eligible in pursuing their higher education learning after finishing the secondary schools. From there, students have choices on how to pursuing their studies, either to join higher learning institutions such as polytechnic, public universities, private universities, community college or university college to enroll the diploma or certificate level. On the other hand, students also may undertake the upper secondary program for two years in school and sit for Sijil Pelajaran Malaysia (SPM).

Students' performance plays major role in determining the quality of our education system. Sijil Pelajaran Malaysia (SPM) is a public examination compulsory to be taken by Form 5 students in Malaysia. The performance gap is not only a school and classroom issue but also a national issue that must be addressed properly [1]. As young generation, their performance in school is important to determine which schools that these children are streamed to further their studies either to daily school, boarding school, semi boarding school or religious school. In effect, this will influence their career path in the future. Factors such as gender [2,3], attendance [1,4], co-curricular activities [5] and family background [6] may influence their performance.

Generally, this study aims to integrate, clean and used that data for automated analysis through data mining techniques. Using data mining techniques is one of the processes of transferring raw data from current educational system to meaningful information that can be used to help the school community to make a right decision to achieve much better results. This proved DM provides means to assist both educators and students, and improve the quality of education. Unfortunately, using the traditional method not only increase the teaching load of the teachers, but also gives unnecessary burdens to students [7]. The result and findings in the study could be used by the management to make faster and more efficient decision in order to map or plan efficient teaching approach for students in the future.

## 2. PROBLEM STATEMENTS

Currently, most schools in Malaysia use Sistem Maklumat Murid (SMM) to collect their information related to family background, income and others as shown in Fig 1. The system gathered information including their names, birth certificate numbers, gender, age, parent's name, parent's job, parent's income, guidance status and siblings. In order to get detail information about students, it is important to access to 'Borang Maklumat Murid' (BMM).

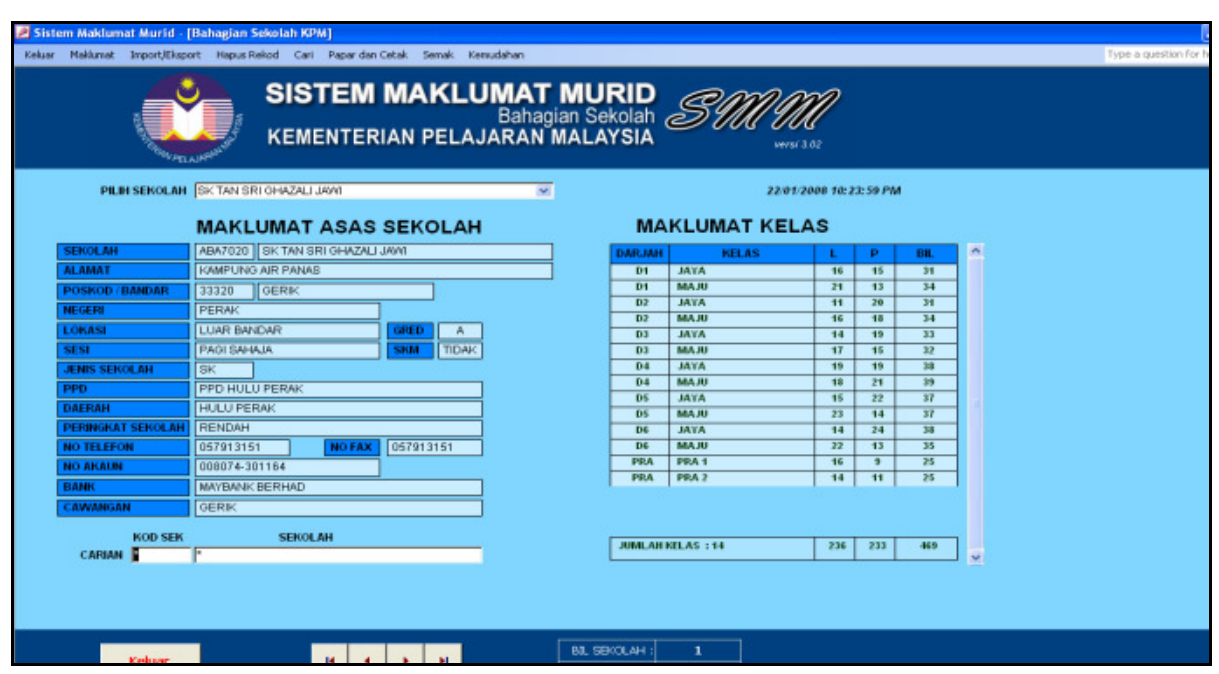

**FIGURE 1**: Interface of SMM

Monthly examination marks are normally keyed-in by respective teacher and stored in Microsoft Excel. Analyses such as crosstabulation and prediction model development have not been explored since data has been key-in independently. School in general is rich with data which is beneficial if the later could be used to help teacher and management understand more about their student background based on their performance. Due to lack of effort in integrating table or database between SMM and student result, this study attempts to uncover the hidden information within SMM data and student result. This study also takes initiative to assist teacher upload their data file in the server for integration and analysis purposes will be done automatically by using automated web based with data mining facilities.

## 3. OBJECTIVE OF STUDY

Generally, the main objectives of this study are to perform integration, cleaning and automated analysis on school data management by using data mining approaches. The specific objectives are listed as:

i. To integrate databases from different sources.
ii. To preprocess data prior to mining process.
iii. To design and implement the prototype of automated data integration.
iv. To evaluate integrated data mining using data mining methods.

## 4. METHODOLOGY

The process flow of the study is illustrated as shown in Fig. 2 that consists of stage Integration, Extract, Cleaning.

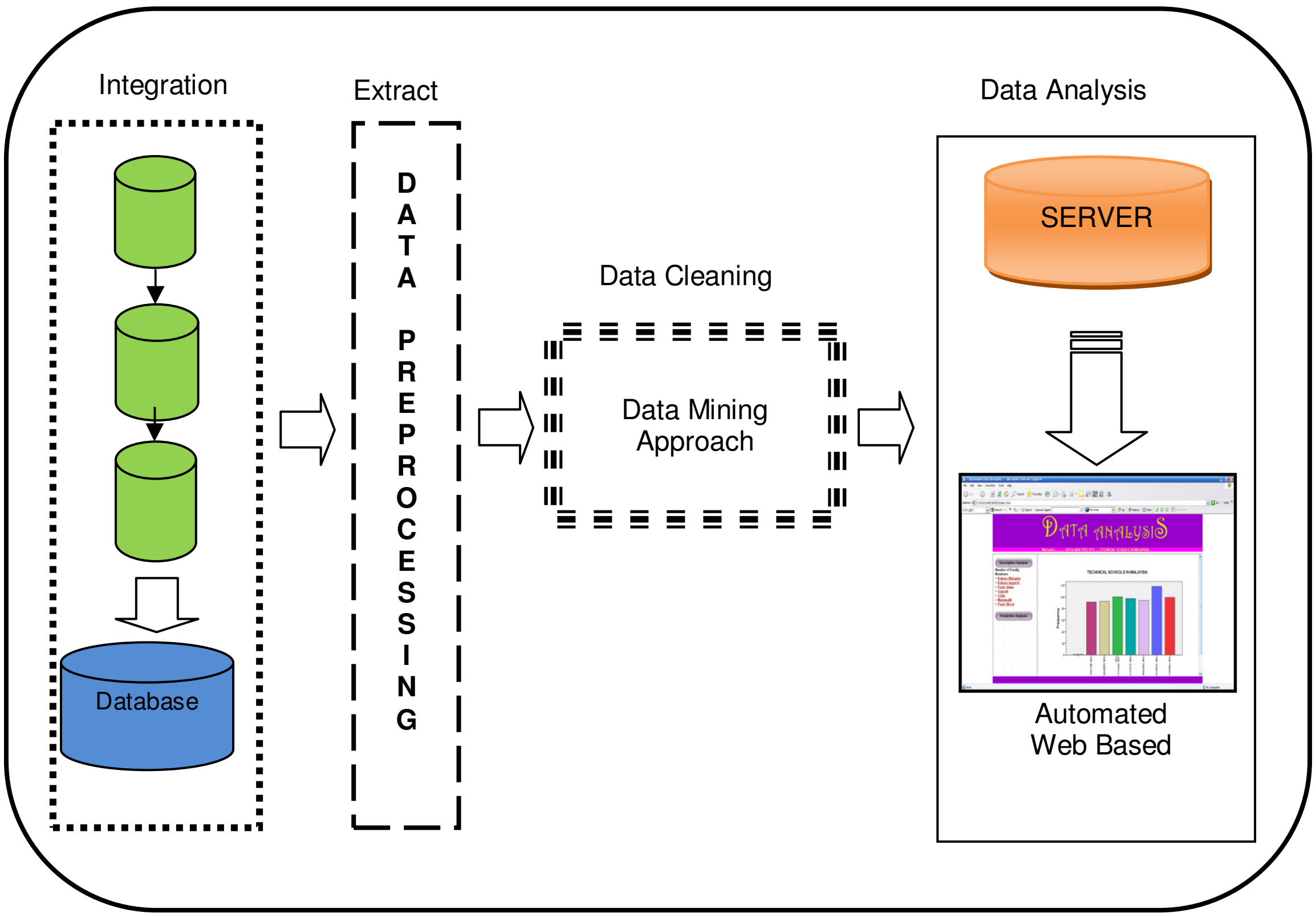

**FIGURE 2**: Process Flow of the study

### 4.1 Data Integration

The initial phase is concerned with collection of data in Microsoft Excel format that integrate seven technical schools in Malaysia. These technical schools include SMK BELAGA SARAWAK, SMK INDERAPURA PAHANG, SMK KEPALA BATAS KEDAH, SMK KUALA KETIL KEDAH, SMK MARANG TERENGGANU, SMK SAMA GAGAH PULAU PINANG, and SMK TENGKU IDRIS SELANGOR. The data will be collected manually and was assist by Jabatan Pelajaran Negeri of each state. These technical schools were show in table 1 below. To analyze the results, descriptive statistics such as frequencies, cross tabulation, and charts were used to describe the output.

| TECHNICAL SCHOOLS | STATE |
|---|---|
| SMK BELAGA | SARAWAK |
| SMK INDERAPURA | PAHANG |
| SMK KEPALA BATAS | KEDAH |
| SMK KUALA KETIL | KEDAH |
| SMK MARANG | TERENGGANU |
| SMK SAMA GAGAH | PULAU PINANG |
| SMK TENGKU IDRIS | SELANGOR |

**Table 1**: Technical Schools that was selected to be samples

### 4.2 Instrument

To make integration of that data, software SPSS 16.0 has been used to combine seven technical schools. The snapshot of the integration process is shown in Fig. 3.

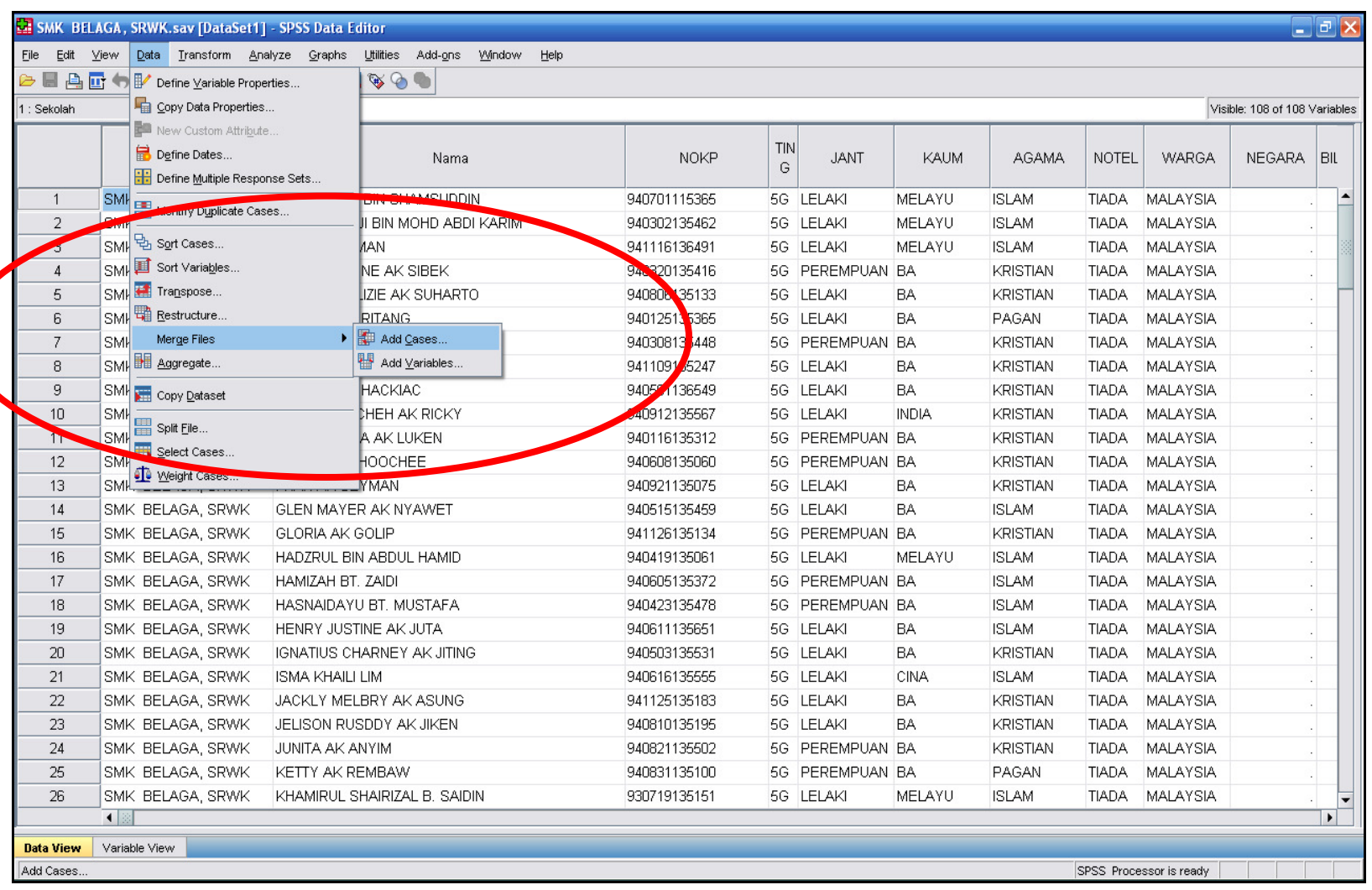

**FIGURE 3** : Snapshot of the integration process

### 4.3 Respondents Data

There are seven technical schools in Malaysia that has been collected to represent as respondent data for this study which is SMK BELAGA SARAWAK (A), SMK INDERAPURA PAHANG (B), SMK KEPALA BATAS KEDAH (C), SMK KUALA KETIL KEDAH (D), SMK MARANG TERENGGANU (E), SMK SAMA GAGAH PULAU PINANG (F), and SMK TENGKU IDRIS SELANGOR (G) and all come in one format that is Microsoft Excel. These data must be integrated together by using SPSS 16.0 software and the flow of process was show in Fig 4 below.

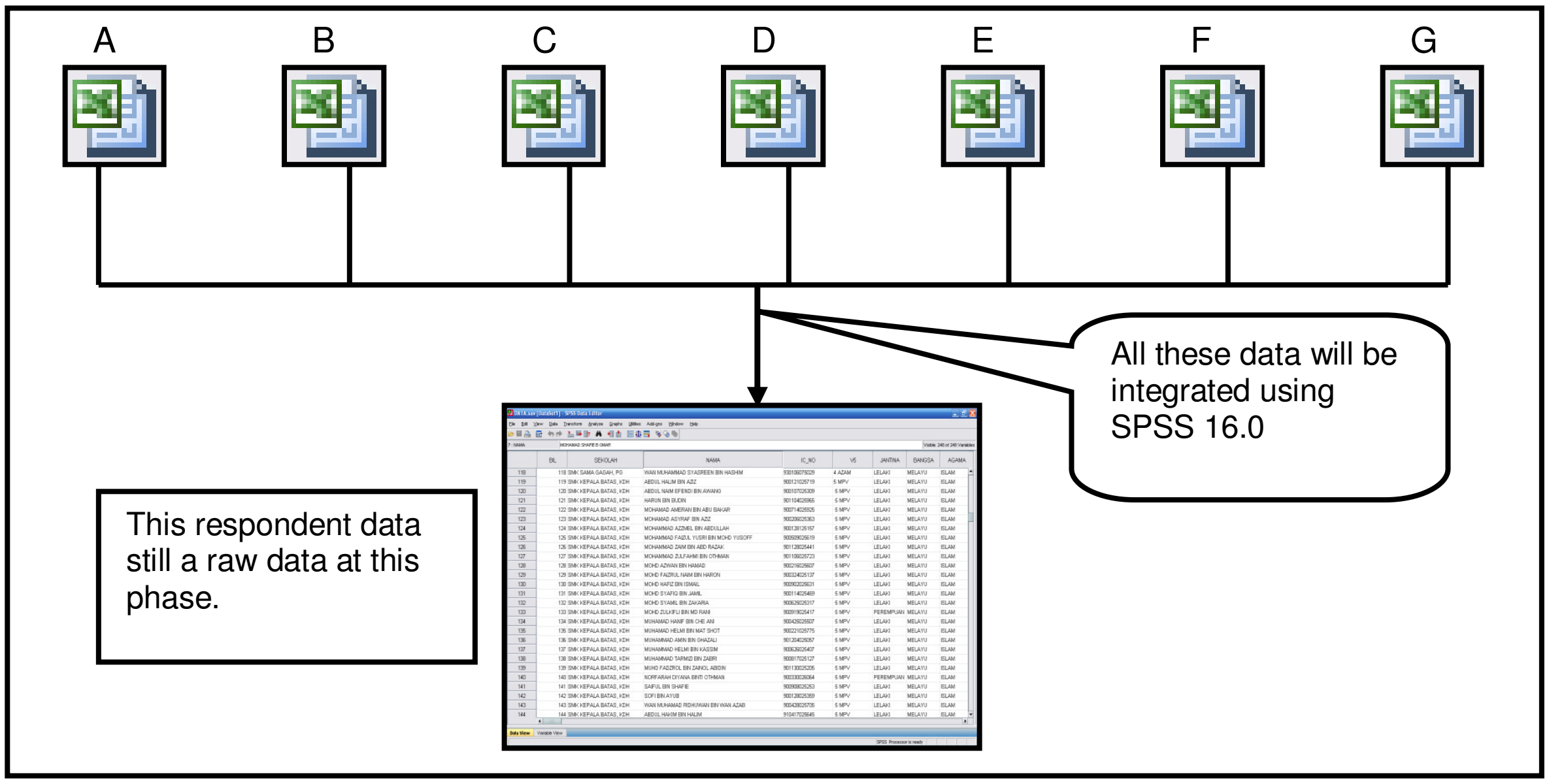

**FIGURE 4**: Flow of Integration in SPSS 16.0 Software

After the integration process, this respondent data are included all technical schools together and 691 respondent have successfully obtained. Before this samples will be process to the next phase which is data cleaning process, this data of all technical schools are show in Table 2.

| Technical Schools | Frequency | % |
|---|---|---|
| SMK BELAGA, SRWK | 91 | 13.2 |
| SMK INDERAPURA, PHG | 92 | 13.3 |
| SMK KEPALA BATAS, KDH | 100 | 14.5 |
| SMK KUALA KETIL, KDH | 97 | 14.0 |
| SMK MARANG, TRG | 94 | 13.6 |
| SMK SAMA GAGAH, PG | 118 | 17.1 |
| SMK TENGKU IDRIS, SLGR | 99 | 14.3 |
| **Total** | **691** | **100.0** |

**TABLE 2**: Respondent Data

### 4.4 Target

According to Shmueli [8], the target dependent variable can be denoted as one of the attributes being used to predict in supervised learning. In this study, SPM attribute's grade is used as the target. SPM attribute's grade was categorized into 5 groups as shown in Table 3.

| SPM Grade | Class |
|---|---|
| 1A ~ 2A | 1 |
| 3B ~ 4B | 2 |
| 5C ~ 6C | 3 |
| 7D ~ 8E | 4 |
| 9G | 5 |

**TABLE 3:** SPM Grade categorized

### 4.5 Data Cleaning

The second phase which is extract and cleaning will be run together and will used data mining approach. Database that is stored in first phase maybe will have high probability of 'dirty data'. Data will be extracted from database and its need to do data preprocesses and then does cleaning. Data cleaning approach should satisfy several requirements. This process was carried out using SPSS version 16.0. Data selection was carried out in this phase and its purpose is to ensure that predicting model can produce more accurate results. Table 4 shows target and various attributes selected before converted to the purposes analysis.

| Type | Input Variable | Domain |
|---|---|---|
| Target | SPM subjects | 1A,2A,3B,4B,5C,6C,7D,8E,9G |
| Attributes | Number of family members | Numeric |
| Attributes | Number of family member still learning | Numeric |
| Attributes | Number of family member receive SPBT | Numeric |
| Attributes | Family Income | RM 500, RM 1200, RM 1900 |
| Attributes | SPBT | YA, TIDAK |

**TABLE 4**: The selected attributes before converted to numeric

Table 5 shows target and various attributes selected after converted for this analysis process. Then, the raw dataset will be changing into numerical forms for experiment sake.

| Type | Input Variable | Domain |
|---|---|---|
| Target | SPM subjects | 1, 2, 3, 4, 5 |
| Attributes | Number of family members | 0, 1, 2, 3, 4, 5, 6, 7, 8, 9, 10, 11, 12, 13 |
| Attributes | Number of family member still learning | 0, 1, 2, 3, 4, 5, 6, 7 |
| Attributes | Number of family member receive SPBT | 0, 1, 2, 3, 4, 5, 6 7 |
| Attributes | Total Income | 1, 2, 3, 4, 5, 6 |
| Attributes | SPBT | 1, 2 |

**TABLE 5**: The selected attributes after converted to numeric

Meanwhile, the process of data transformation was done to ensure the data formats are in appropriate form that can using by SPSS 16.0 tool. (See Fig. 5 to Fig. 8)

*Untitled2 [DataSet1] - SPSS Data Editor

1 : NAMASEKOLAH SMK SAMA GAGAH, PG

| | NAMASEKOLAH | NAMACALON | NOKP | TING | JANT | KAUM | AGAMA |
|---|---|---|---|---|---|---|---|
| 1 | SMK SAMA GAGAH, PG | MOHAMAD SHALAHUDIN B ABD RAHMAN | 880207355553 | 5 AZAM | LELAKI | MELAYU | ISLAM |
| 2 | SMK SAMA GAGAH, PG | MOHAMAD SYAFIE B ABDUL AZIZ | 880729075161 | 5 AZAM | LELAKI | MELAYU | ISLAM |
| 3 | SMK SAMA GAGAH, PG | MOHAMMAD AZRUL B MOHD RAHIM | 880811355429 | 5 AZAM | LELAKI | MELAYU | ISLAM |
| 4 | SMK SAMA GAGAH, PG | MOHAMMAD SHOPIAN B KASIM | 880723355605 | 5 AZAM | LELAKI | MELAYU | ISLAM |
| 5 | SMK SAMA GAGAH, PG | MOHD FARID B RODZI | 880519355261 | 5 AZAM | LELAKI | MELAYU | ISLAM |
| 6 | SMK SAMA GAGAH, PG | MOHD FIRDAUS B ZAKARIAH | 880401355637 | 5 AZAM | LELAKI | MELAYU | ISLAM |
| 7 | SMK SAMA GAGAH, PG | MOHAMAD SHAFIE B OMAR | 880521355535 | 5 AZAM | LELAKI | MELAYU | ISLAM |
| 8 | SMK SAMA GAGAH, PG | MOHD SHAHROL AMRIN B SHAHIDAN | 880131025717 | 5 AZAM | LELAKI | MELAYU | ISLAM |
| 9 | SMK SAMA GAGAH, PG | MOHD TAHIR B MD DESA | 881003355345 | 5 AZAM | LELAKI | MELAYU | ISLAM |
| 10 | SMK SAMA GAGAH, PG | MUHAMAD FAUZI B MOHD NOR | 880728075063 | 5 AZAM | LELAKI | MELAYU | ISLAM |
| 11 | SMK SAMA GAGAH, PG | MUHAMAD RAZIF B ABD KADIR | 880701355233 | 5 AZAM | LELAKI | MELAYU | ISLAM |
| 12 | SMK SAMA GAGAH, PG | MUHAMMAD AMIRUDDIN B ISMAIL | 880401355653 | 5 AZAM | LELAKI | MELAYU | ISLAM |
| 13 | SMK SAMA GAGAH, PG | MUHAMMAD NAAIN B ABD RASHID | 880824355209 | 5 AZAM | LELAKI | MELAYU | ISLAM |
| 14 | SMK SAMA GAGAH, PG | MUHAMMAD SYAHFAIZ B MAHAD ALI | 880225355063 | 5 AZAM | LELAKI | MELAYU | ISLAM |
| 15 | SMK SAMA GAGAH, PG | MUHAMMAD ZAINI B CHE AHMAD | 880608355335 | 5 AZAM | LELAKI | MELAYU | ISLAM |
| 16 | SMK SAMA GAGAH, PG | NASRUL FAMI B ROSLAN | 880411355171 | 5 AZAM | LELAKI | MELAYU | ISLAM |
| 17 | SMK SAMA GAGAH, PG | SHAHRIL AIZAT B NORDIN | 880528355343 | 5 AZAM | LELAKI | MELAYU | ISLAM |
| 18 | SMK SAMA GAGAH, PG | WAN MOHAMMAD SYAMIL B HASHIM | 880627086767 | 5 AZAM | LELAKI | MELAYU | ISLAM |
| 19 | SMK SAMA GAGAH, PG | YUVANESWARAN A/L SIVALINGAM | 880723075377 | 5 AZAM | LELAKI | INDIAN | HINDU |
| 20 | SMK SAMA GAGAH, PG | ABDUL MUIZ B ARIFFIN | 890707075063 | 5 AZAM | LELAKI | MELAYU | ISLAM |
| 21 | SMK SAMA GAGAH, PG | CHEW HUANG KEAT | 880717355203 | 5 AZAM | LELAKI | CINA | BUDDHA |

**FIGURE 5**: Sample of raw dataset before select attributes and converted using SPSS 16.0

DATA.sav [DataSet1] - SPSS Data Editor

1 : SCHOOLS SMK SAMA GAGAH, PG

| | SCHOOLS | RESPONDENT_NAME | NUM_FAMILY_MEMBERS | NUM_FAMILY_MEMBERS_LEARN | NUM_FAMILY_MEMBERS_RECEIVE_SPBT | FAMILY_INCOME | SPBT | |
|---|---|---|---|---|---|---|---|---|
| 1 | SMK SAMA GAGAH, PG | MOHAMAD SHALAHUDIN B ABD RAHMAN | 4 | 2 | 2 | 1800 | YA | 6C |
| 2 | SMK SAMA GAGAH, PG | MOHAMAD SYAFIE B ABDUL AZIZ | 3 | 1 | 1 | 1000 | YA | 7D |
| 3 | SMK SAMA GAGAH, PG | MOHAMMAD AZRUL B MOHD RAHIM | 5 | 2 | 2 | 1500 | YA | 8E |
| 4 | SMK SAMA GAGAH, PG | MOHAMMAD SHOPIAN B KASIM | 6 | 3 | 3 | 1000 | YA | 7D |
| 5 | SMK SAMA GAGAH, PG | MOHD FARID B RODZI | 6 | 1 | 1 | 1200 | YA | 8E |
| 6 | SMK SAMA GAGAH, PG | MOHD FIRDAUS B ZAKARIAH | 4 | 2 | 2 | 1000 | YA | 9G |
| 7 | SMK SAMA GAGAH, PG | MOHAMAD SHAFIE B OMAR | 5 | 3 | 3 | 500 | YA | 6C |
| 8 | SMK SAMA GAGAH, PG | MOHD SHAHROL AMRIN B SHAHIDAN | 6 | 3 | 3 | 750 | YA | 9G |
| 9 | SMK SAMA GAGAH, PG | MOHD TAHIR B MD DESA | 3 | 1 | 1 | 650 | YA | 8E |
| 10 | SMK SAMA GAGAH, PG | MUHAMAD FAUZI B MOHD NOR | 4 | 2 | 2 | 750 | YA | 7D |
| 11 | SMK SAMA GAGAH, PG | MUHAMAD RAZIF B ABD KADIR | 5 | 2 | 2 | 500 | YA | 7D |
| 12 | SMK SAMA GAGAH, PG | MUHAMMAD AMIRUDDIN B ISMAIL | 6 | 3 | 3 | 750 | YA | 7D |
| 13 | SMK SAMA GAGAH, PG | MUHAMMAD NAAIN B ABD RASHID | 5 | 2 | 2 | 700 | YA | 7D |
| 14 | SMK SAMA GAGAH, PG | MUHAMMAD SYAHFAIZ B MAHAD ALI | 4 | 2 | 2 | 1350 | YA | 9G |
| 15 | SMK SAMA GAGAH, PG | MUHAMMAD ZAINI B CHE AHMAD | 6 | 3 | 3 | 650 | YA | 8E |
| 16 | SMK SAMA GAGAH, PG | NASRUL FAMI B ROSLAN | 3 | 1 | 1 | 700 | YA | 5C |
| 17 | SMK SAMA GAGAH, PG | SHAHRIL AIZAT B NORDIN | 4 | 2 | 2 | 700 | YA | 7D |
| 18 | SMK SAMA GAGAH, PG | WAN MOHAMMAD SYAMIL B HASHIM | 6 | 3 | 3 | 500 | YA | 7D |
| 19 | SMK SAMA GAGAH, PG | YUVANESWARAN A/L SIVALINGAM | 3 | 1 | 1 | 1000 | YA | 7D |
| 20 | SMK SAMA GAGAH, PG | ABDUL MUIZ B ARIFFIN | 5 | 2 | 2 | 3000 | YA | 8E |
| 21 | SMK SAMA GAGAH, PG | CHEW HUANG KEAT | 3 | 1 | 1 | 750 | YA | 9G |
| 22 | SMK SAMA GAGAH, PG | CHU WEN JIAN | 4 | 2 | 2 | 450 | YA | 9G |

**FIGURE 6**: The selected attributes

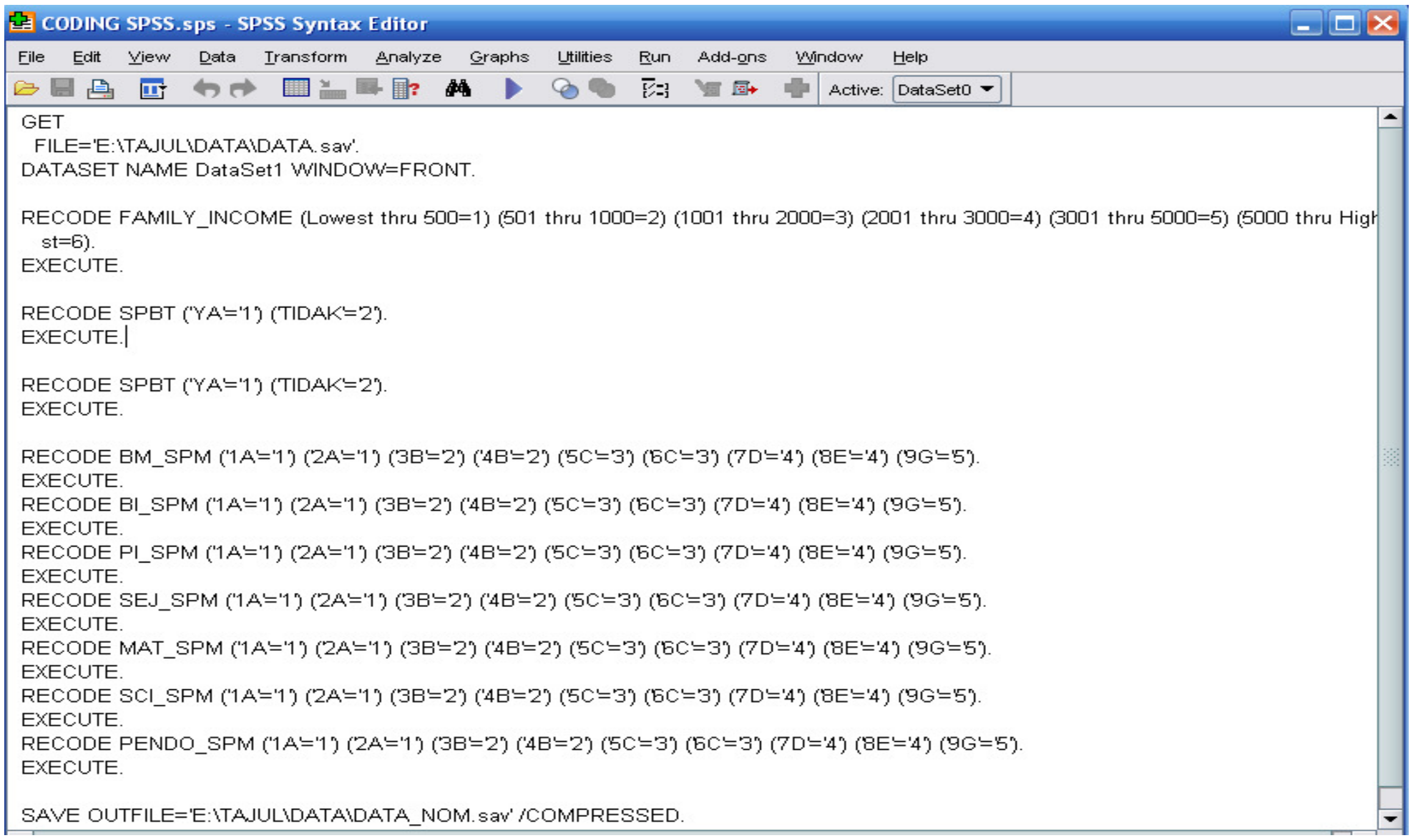

**FIGURE 7**: Process of data set conversion using Syntax SPSS in SPSS 16.0

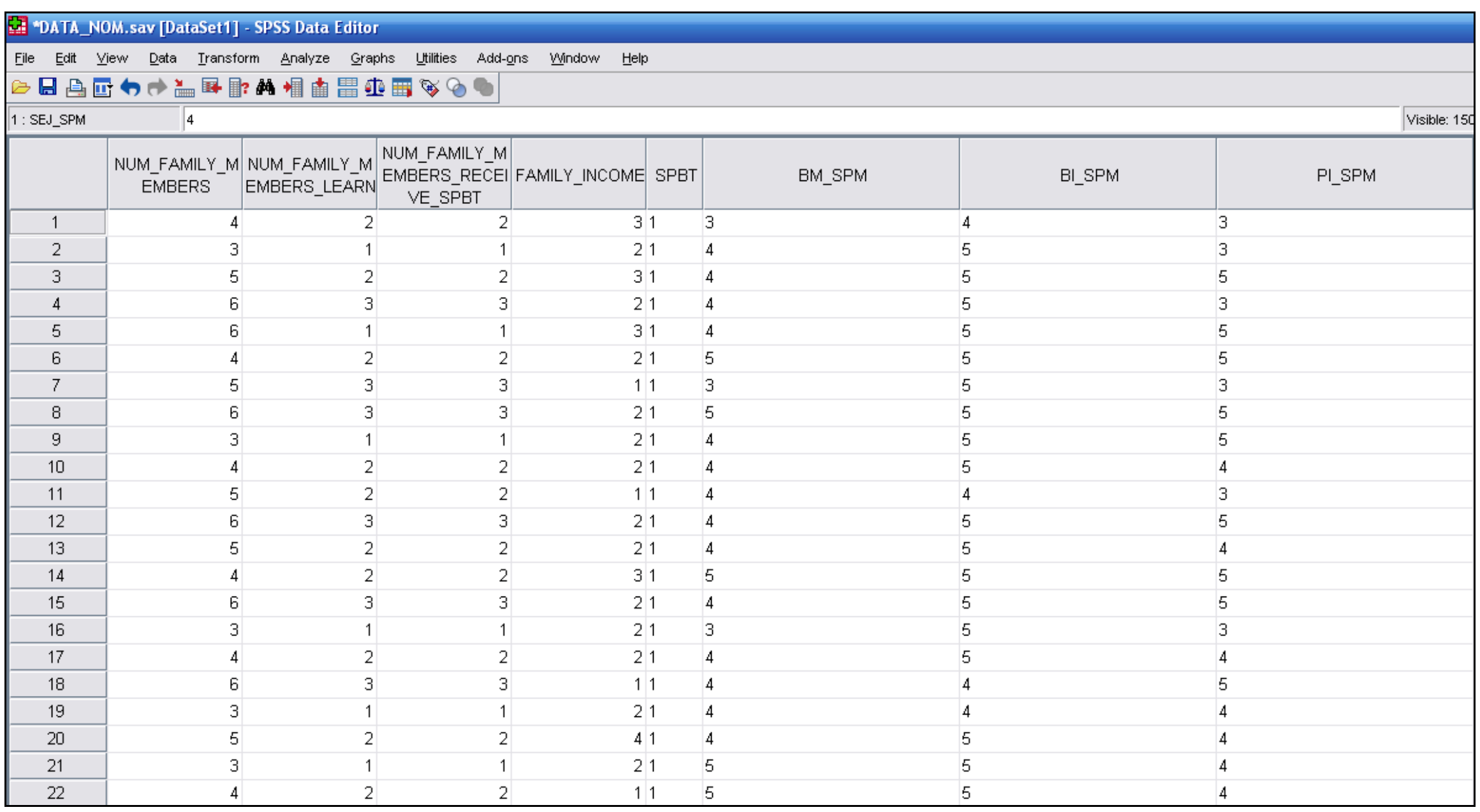

| | NUM_FAMILY_MEMBERS | NUM_FAMILY_MEMBERS_LEARN | NUM_FAMILY_MEMBERS_RECEIVE_SPBT | FAMILY_INCOME | SPBT | BM_SPM | BI_SPM | PI_SPM |
|---|---|---|---|---|---|---|---|---|
| 1 | 4 | 2 | 2 | 3 | 1 | 3 | 4 | 3 |
| 2 | 3 | 1 | 1 | 2 | 1 | 4 | 5 | 3 |
| 3 | 5 | 2 | 2 | 3 | 1 | 4 | 5 | 5 |
| 4 | 6 | 3 | 3 | 2 | 1 | 4 | 5 | 3 |
| 5 | 6 | 1 | 1 | 3 | 1 | 4 | 5 | 5 |
| 6 | 4 | 2 | 2 | 2 | 1 | 5 | 5 | 5 |
| 7 | 5 | 3 | 3 | 1 | 1 | 3 | 5 | 3 |
| 8 | 6 | 3 | 3 | 2 | 1 | 5 | 5 | 5 |
| 9 | 3 | 1 | 1 | 2 | 1 | 4 | 5 | 5 |
| 10 | 4 | 2 | 2 | 2 | 1 | 4 | 5 | 4 |
| 11 | 5 | 2 | 2 | 1 | 1 | 4 | 4 | 3 |
| 12 | 6 | 3 | 3 | 2 | 1 | 4 | 5 | 5 |
| 13 | 5 | 2 | 2 | 2 | 1 | 4 | 5 | 4 |
| 14 | 4 | 2 | 2 | 3 | 1 | 5 | 5 | 5 |
| 15 | 6 | 3 | 3 | 2 | 1 | 4 | 5 | 5 |
| 16 | 3 | 1 | 1 | 2 | 1 | 3 | 5 | 3 |
| 17 | 4 | 2 | 2 | 2 | 1 | 4 | 5 | 4 |
| 18 | 6 | 3 | 3 | 1 | 1 | 4 | 4 | 5 |
| 19 | 3 | 1 | 1 | 2 | 1 | 4 | 4 | 4 |
| 20 | 5 | 2 | 2 | 4 | 1 | 4 | 5 | 4 |
| 21 | 3 | 1 | 1 | 2 | 1 | 5 | 5 | 4 |
| 22 | 4 | 2 | 2 | 1 | 1 | 5 | 5 | 4 |

**FIGURE 8**: Data set after converted in SPSS 16.0

## 5. RESULT AND FINDING

This section will illustrate the analysis and result gained from the data collected. The descriptive analysis has been carried out to get some information from the outcomes of each prediction variable aligned with the targeted output.

### 5.1 Data Integration

This study has illustrate how the process of data integration on data set has been applied for make it for the data analysis part. As you know, if the process of data integration has some error or fail, it will lead to the wrong analysis and wrong result. This phase is so important and it will be used SPSS 16.0 (Syntax) instrument to make this process done. There are seven data set which from technical schools in Malaysia has been selected as respondent data and need to integrated all together include SMK BELAGA SARAWAK, SMK INDERAPURA PAHANG, SMK KEPALA BATAS KEDAH, SMK KUALA KETIL KEDAH, SMK MARANG TERENGGANU, SMK SAMA GAGAH PULAU PINANG, and SMK TENGKU IDRIS SELANGOR. SPSS 16.0 software can be used manually to integrated these data but for this study purpose, it will be used Syntax editor in SPSS 16.0 to make it automated integration for these data. The original sources of dataset are in Microsoft Excel format and it will be imported to SPSS 16.0 program by using Syntax editor. The process of integration will be show below.

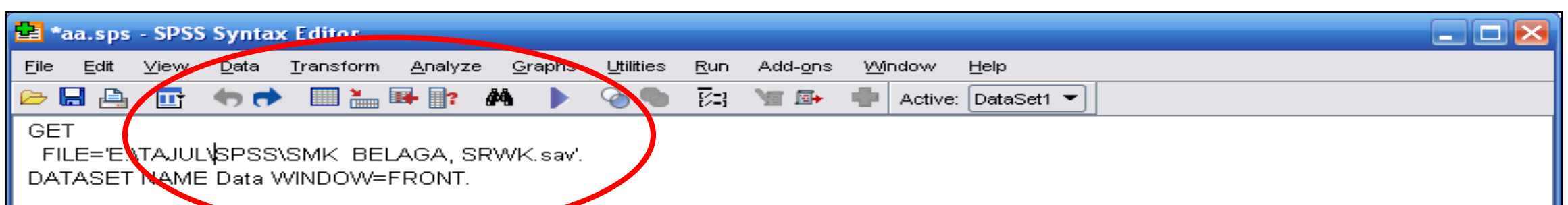

**FIGURE 9**: Source code in syntax editor to import data into SPSS 16.0

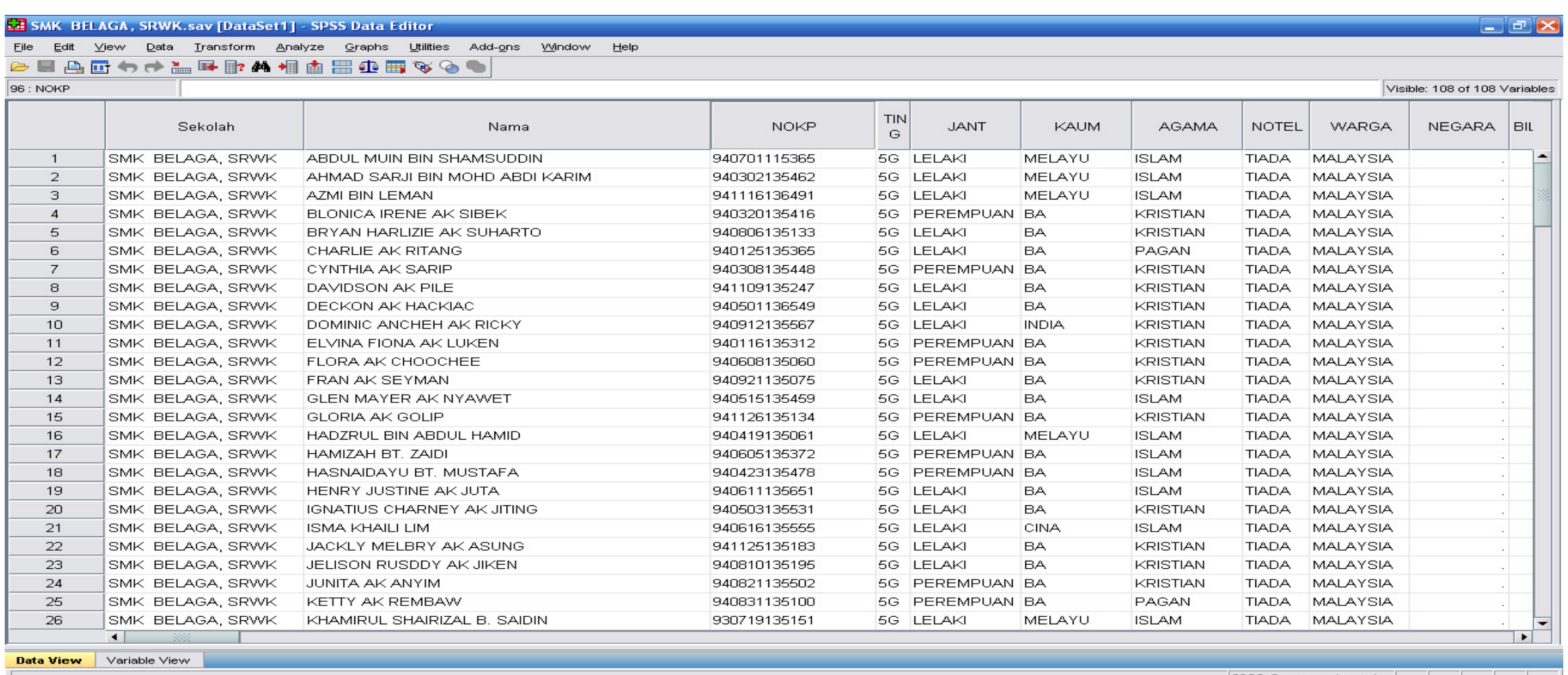

| | Sekolah | Nama | NOKP | TING | JANT | KAUM | AGAMA | NOTEL | WARGA | NEGARA | BIL |
|---|---|---|---|---|---|---|---|---|---|---|---|
| 1 | SMK BELAGA, SRWK | ABDUL MUIN BIN SHAMSUDDIN | 940701115365 | 5G | LELAKI | MELAYU | ISLAM | TIADA | MALAYSIA | . | |
| 2 | SMK BELAGA, SRWK | AHMAD SARJI BIN MOHD ABDI KARIM | 940302135462 | 5G | LELAKI | MELAYU | ISLAM | TIADA | MALAYSIA | . | |
| 3 | SMK BELAGA, SRWK | AZMI BIN LEMAN | 941116136491 | 5G | LELAKI | MELAYU | ISLAM | TIADA | MALAYSIA | . | |
| 4 | SMK BELAGA, SRWK | BLONICA IRENE AK SIBEK | 940320135416 | 5G | PEREMPUAN | BA | KRISTIAN | TIADA | MALAYSIA | . | |
| 5 | SMK BELAGA, SRWK | BRYAN HARLIZIE AK SUHARTO | 940806135133 | 5G | LELAKI | BA | KRISTIAN | TIADA | MALAYSIA | . | |
| 6 | SMK BELAGA, SRWK | CHARLIE AK RITANG | 940125135365 | 5G | LELAKI | BA | PAGAN | TIADA | MALAYSIA | . | |
| 7 | SMK BELAGA, SRWK | CYNTHIA AK SARIP | 940308135448 | 5G | PEREMPUAN | BA | KRISTIAN | TIADA | MALAYSIA | . | |
| 8 | SMK BELAGA, SRWK | DAVIDSON AK PILE | 941109135247 | 5G | LELAKI | BA | KRISTIAN | TIADA | MALAYSIA | . | |
| 9 | SMK BELAGA, SRWK | DECKON AK HACKIAC | 940501136549 | 5G | LELAKI | BA | KRISTIAN | TIADA | MALAYSIA | . | |
| 10 | SMK BELAGA, SRWK | DOMINIC ANCHEH AK RICKY | 940912135567 | 5G | LELAKI | INDIA | KRISTIAN | TIADA | MALAYSIA | . | |
| 11 | SMK BELAGA, SRWK | ELVINA FIONA AK LUKEN | 940116135312 | 5G | PEREMPUAN | BA | KRISTIAN | TIADA | MALAYSIA | . | |
| 12 | SMK BELAGA, SRWK | FLORA AK CHOOCHEE | 940608135060 | 5G | PEREMPUAN | BA | KRISTIAN | TIADA | MALAYSIA | . | |
| 13 | SMK BELAGA, SRWK | FRAN AK SEYMAN | 940921135075 | 5G | LELAKI | BA | KRISTIAN | TIADA | MALAYSIA | . | |
| 14 | SMK BELAGA, SRWK | GLEN MAYER AK NYAWET | 940515135459 | 5G | LELAKI | BA | ISLAM | TIADA | MALAYSIA | . | |
| 15 | SMK BELAGA, SRWK | GLORIA AK GOLIP | 941126135134 | 5G | PEREMPUAN | BA | KRISTIAN | TIADA | MALAYSIA | . | |
| 16 | SMK BELAGA, SRWK | HADZRUL BIN ABDUL HAMID | 940419135061 | 5G | LELAKI | MELAYU | ISLAM | TIADA | MALAYSIA | . | |
| 17 | SMK BELAGA, SRWK | HAMIZAH BT. ZAIDI | 940605135372 | 5G | PEREMPUAN | BA | ISLAM | TIADA | MALAYSIA | . | |
| 18 | SMK BELAGA, SRWK | HASNAIDAYU BT. MUSTAFA | 940423135478 | 5G | PEREMPUAN | BA | ISLAM | TIADA | MALAYSIA | . | |
| 19 | SMK BELAGA, SRWK | HENRY JUSTINE AK JUTA | 940611135651 | 5G | LELAKI | BA | ISLAM | TIADA | MALAYSIA | . | |
| 20 | SMK BELAGA, SRWK | IGNATIUS CHARNEY AK JITING | 940503135531 | 5G | LELAKI | BA | KRISTIAN | TIADA | MALAYSIA | . | |
| 21 | SMK BELAGA, SRWK | ISMA KHAILI LIM | 940616135555 | 5G | LELAKI | CINA | ISLAM | TIADA | MALAYSIA | . | |
| 22 | SMK BELAGA, SRWK | JACKLY MELBRY AK ASUNG | 941125135183 | 5G | LELAKI | BA | KRISTIAN | TIADA | MALAYSIA | . | |
| 23 | SMK BELAGA, SRWK | JELISON RUSDDY AK JIKEN | 940810135195 | 5G | LELAKI | BA | KRISTIAN | TIADA | MALAYSIA | . | |
| 24 | SMK BELAGA, SRWK | JUNITA AK ANYIM | 940821135502 | 5G | PEREMPUAN | BA | KRISTIAN | TIADA | MALAYSIA | . | |
| 25 | SMK BELAGA, SRWK | KETTY AK REMBAW | 940831135100 | 5G | PEREMPUAN | BA | PAGAN | TIADA | MALAYSIA | . | |
| 26 | SMK BELAGA, SRWK | KHAMIRUL SHAIRIZAL B. SAIDIN | 930719135151 | 5G | LELAKI | MELAYU | ISLAM | TIADA | MALAYSIA | . | |

**FIGURE 10**: Dataset in SPSS 16.0 after import the data using syntax editor

Fig. 9 and 10 above have show how the dataset from Microsoft Excel format will be imported using SPSS syntax and will produce the output in SPSS 16.0 format by automatically without need used the manual from SPSS .

**SYNTAX EDITOR :**

```
GET
  FILE='E:\TAJUL\SPSS\SMK  BELAGA, SRWK.sav'.
DATASET NAME Data WINDOW=FRONT.

MATCH FILES /FILE=*
  /FILE='E:\TAJUL\SPSS\SMK  INDERAPURA, PHG.sav'
  /RENAME (AGAMA ALAMAT AM BANDAR BI BI_A BI_B BI_C BI_D BI_E BI_F BI_G BI_H BIASISWA
    BILADIKBERADIKTERIMARMT BILADIKBERADIKTERIMASPBT
BILADIKBERADIKTINGGALDIASRAMA BILISIKELUARGA
    BILYANGMASIHBELAJAR BM BM_A BM_B BM_C BM_D BM_E BM_F BM_G BM_H JANT
JARAKSEKOLAH JENISASRAMA
    JENISBIASISWA JUMLAHPENDAPATAN KAUM MAT MAT_A MAT_B MAT_C MAT_D MAT_E
MAT_F MAT_G MAT_H Nama
    NAMAWARIS NEGARA NEGERI NOKP NOKPWARIS NOTEL PAKAIANSERAGAM PEKERJAAN
PEKERJAAN_A PEKERJAAN_B
    PENDAPATANBAPA PENDAPATANIBU PENDAPATANPENJAGA PENDO PENDO_A PENDO_B
PENDO_C PENDO_D PENDO_E
    PENDO_F PENDO_G PENDO_H PERKAPITA PERALATANSEKOLAH PI PI_A PI_B PI_C PI_D
PI_E PI_F PI_G PI_H
    POSKOD PSS RMT SCI SCI_A SCI_B SCI_C SCI_D SCI_E SCI_F SCI_G SCI_H SEJ SEJ_A
SEJ_B SEJ_C SEJ_D
    SEJ_E SEJ_F SEJ_G SEJ_H Sekolah SPBT STATUSMURIDANAKYATIM
STATUSPENJAGAMURID TING TINGGALDIASRAMA
    TUISYEN V108 WARGA YURAN = d0 d1 d2 d3 d4 d5 d6 d7 d8 d9 d10 d11 d12 d13 d14 d15 d16 d17
d18 d19
    d20 d21 d22 d23 d24 d25 d26 d27 d28 d29 d30 d31 d32 d33 d34 d35 d36 d37 d38 d39 d40 d41 d42
d43 d44
    d45 d46 d47 d48 d49 d50 d51 d52 d53 d54 d55 d56 d57 d58 d59 d60 d61 d62 d63 d64 d65 d66 d67
d68 d69
```

**FIGURE 11**: Source code in syntax editor to make integration of all the respondent data

Fig. 11 above is some the source code in syntax SPSS to make integration of all these dataset by automatically control in syntax SPSS. Fig 11 above not include all the source code in syntax SPSS because of the space here. So it just can show four technical schools that will integrate together and the output will be save to another file in SPSS format which is ‘DATA_COMBINE’ as show in Fig. 12 below.

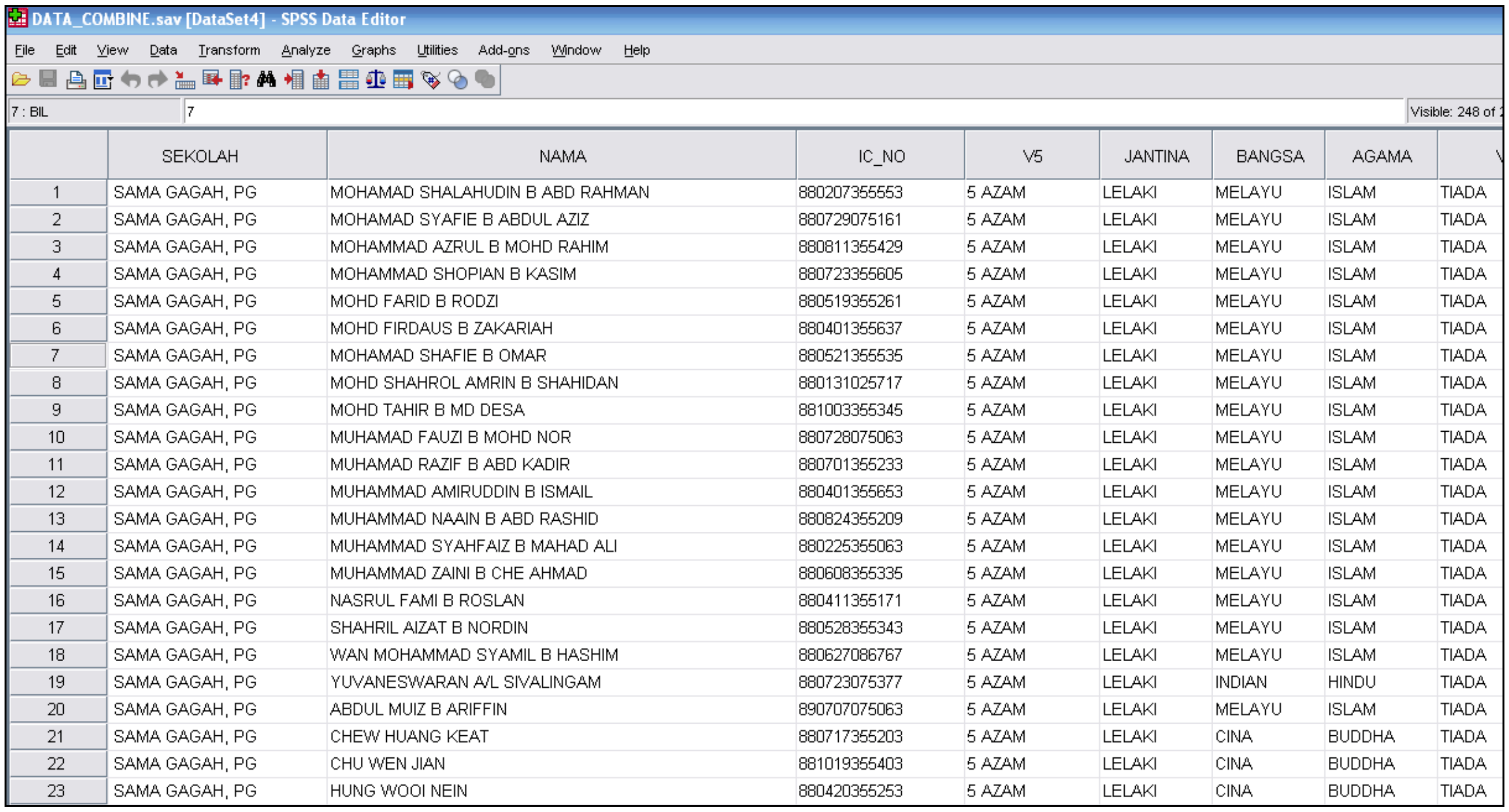

| | SEKOLAH | NAMA | IC_NO | V5 | JANTINA | BANGSA | AGAMA | V |
|---|---|---|---|---|---|---|---|---|
| 1 | SAMA GAGAH, PG | MOHAMAD SHALAHUDIN B ABD RAHMAN | 880207355553 | 5 AZAM | LELAKI | MELAYU | ISLAM | TIADA |
| 2 | SAMA GAGAH, PG | MOHAMAD SYAFIE B ABDUL AZIZ | 880729075161 | 5 AZAM | LELAKI | MELAYU | ISLAM | TIADA |
| 3 | SAMA GAGAH, PG | MOHAMMAD AZRUL B MOHD RAHIM | 880811355429 | 5 AZAM | LELAKI | MELAYU | ISLAM | TIADA |
| 4 | SAMA GAGAH, PG | MOHAMMAD SHOPIAN B KASIM | 880723355605 | 5 AZAM | LELAKI | MELAYU | ISLAM | TIADA |
| 5 | SAMA GAGAH, PG | MOHD FARID B RODZI | 880519355261 | 5 AZAM | LELAKI | MELAYU | ISLAM | TIADA |
| 6 | SAMA GAGAH, PG | MOHD FIRDAUS B ZAKARIAH | 880401355637 | 5 AZAM | LELAKI | MELAYU | ISLAM | TIADA |
| 7 | SAMA GAGAH, PG | MOHAMAD SHAFIE B OMAR | 880521355535 | 5 AZAM | LELAKI | MELAYU | ISLAM | TIADA |
| 8 | SAMA GAGAH, PG | MOHD SHAHROL AMRIN B SHAHIDAN | 880131025717 | 5 AZAM | LELAKI | MELAYU | ISLAM | TIADA |
| 9 | SAMA GAGAH, PG | MOHD TAHIR B MD DESA | 881003355345 | 5 AZAM | LELAKI | MELAYU | ISLAM | TIADA |
| 10 | SAMA GAGAH, PG | MUHAMAD FAUZI B MOHD NOR | 880728075063 | 5 AZAM | LELAKI | MELAYU | ISLAM | TIADA |
| 11 | SAMA GAGAH, PG | MUHAMAD RAZIF B ABD KADIR | 880701355233 | 5 AZAM | LELAKI | MELAYU | ISLAM | TIADA |
| 12 | SAMA GAGAH, PG | MUHAMMAD AMIRUDDIN B ISMAIL | 880401355653 | 5 AZAM | LELAKI | MELAYU | ISLAM | TIADA |
| 13 | SAMA GAGAH, PG | MUHAMMAD NAAIN B ABD RASHID | 880824355209 | 5 AZAM | LELAKI | MELAYU | ISLAM | TIADA |
| 14 | SAMA GAGAH, PG | MUHAMMAD SYAHFAIZ B MAHAD ALI | 880225355063 | 5 AZAM | LELAKI | MELAYU | ISLAM | TIADA |
| 15 | SAMA GAGAH, PG | MUHAMMAD ZAINI B CHE AHMAD | 880608355335 | 5 AZAM | LELAKI | MELAYU | ISLAM | TIADA |
| 16 | SAMA GAGAH, PG | NASRUL FAMI B ROSLAN | 880411355171 | 5 AZAM | LELAKI | MELAYU | ISLAM | TIADA |
| 17 | SAMA GAGAH, PG | SHAHRIL AIZAT B NORDIN | 880528355343 | 5 AZAM | LELAKI | MELAYU | ISLAM | TIADA |
| 18 | SAMA GAGAH, PG | WAN MOHAMMAD SYAMIL B HASHIM | 880627086767 | 5 AZAM | LELAKI | MELAYU | ISLAM | TIADA |
| 19 | SAMA GAGAH, PG | YUVANESWARAN A/L SIVALINGAM | 880723075377 | 5 AZAM | LELAKI | INDIAN | HINDU | TIADA |
| 20 | SAMA GAGAH, PG | ABDUL MUIZ B ARIFFIN | 890707075063 | 5 AZAM | LELAKI | MELAYU | ISLAM | TIADA |
| 21 | SAMA GAGAH, PG | CHEW HUANG KEAT | 880717355203 | 5 AZAM | LELAKI | CINA | BUDDHA | TIADA |
| 22 | SAMA GAGAH, PG | CHU WEN JIAN | 881019355403 | 5 AZAM | LELAKI | CINA | BUDDHA | TIADA |
| 23 | SAMA GAGAH, PG | HUNG WOOI NEIN | 880420355253 | 5 AZAM | LELAKI | CINA | BUDDHA | TIADA |

**FIGURE 12**: Output of data integration using Syntax SPSS.

## 5.2 Data Analysis

Data analysis for this study is involved of descriptive and predictive analysis that will be handling using Syntax SPSS code in SPSS 16.0. All the process in analysis part will be done by automatically using Syntax SPSS.

## 5.3 Descriptive Analysis

The collected data in this study has been processed using Syntax SPSS in SPSS version 16.0 to produce the experimental results. The analysis is divided into five main factors, such as number of family members, number of family member still learning, number of family member receive SPBT, family income and SPBT .

## 5.4 Number Of Family Members

The number of family members is one of the attribute of dataset that will be analyzed to see how it can influence the SPM result of student in technical schools. Initial analysis is done to make sure there is no missing value in the collected data. The analysis is shown below:

**Statistics :**

| NUM_FAMILY_MEMBERS | | |
|---|---|---|
| N | Valid | 691 |
| | Missing | 0 |

**Syntax SPSS Code :**

```
GET
FILE='E:\TAJUL\DATA\DATA_COMBINE.sav'
.
DATASET NAME DataSet1
WINDOW=FRONT.

FREQUENCIES
VARIABLES=NUM_FAMILY_MEMBERS
  /ORDER=ANALYSIS.
```

**TABLE 6**: Frequency data of Number of family members and source code in Syntax SPSS

Table 6 denotes there is no missing value for this attribute for used in the experiment and it was generate using Syntax SPSS code. In this attribute, there are 691 students whose number of family member's distribution is shown in Table 7.

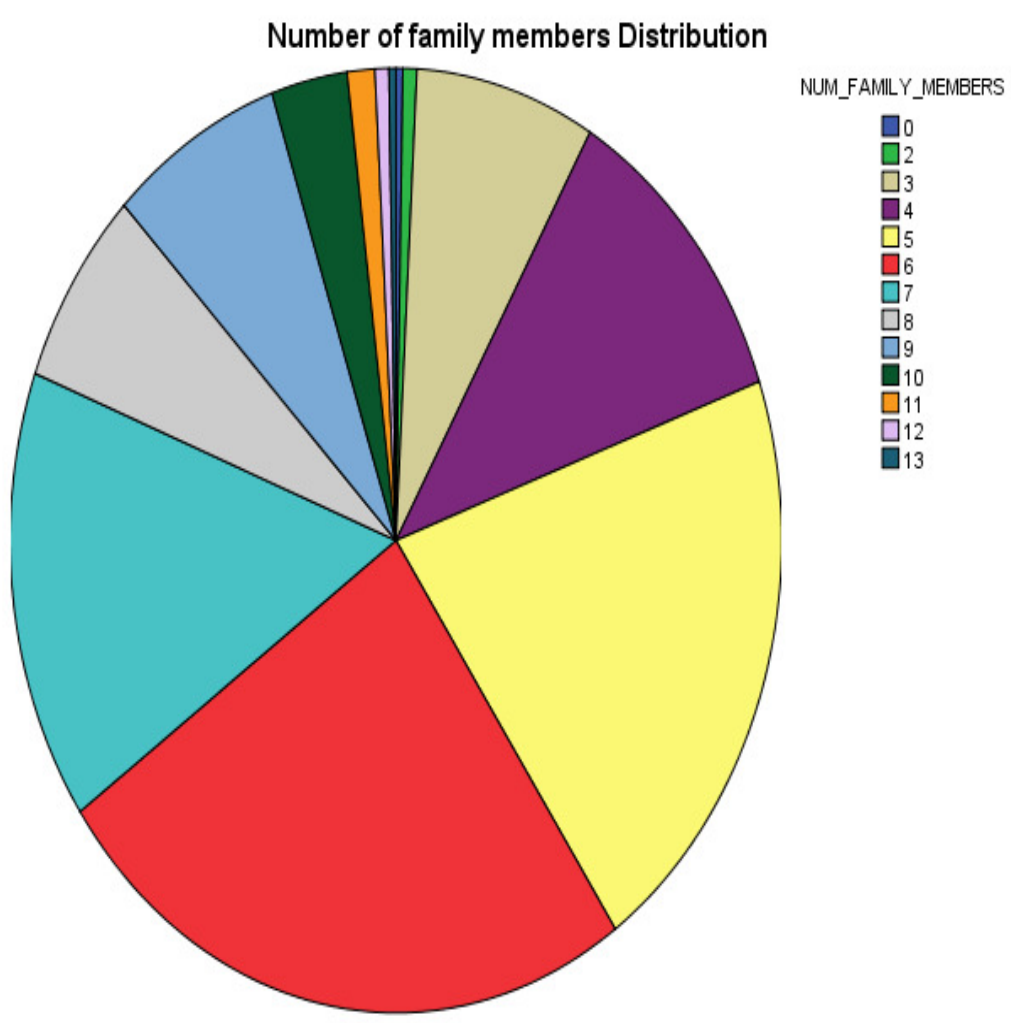

**Syntax SPSS Code:**

```
GGRAPH
  /GRAPHDATASET NAME="graphdataset"
VARIABLES=NUM_FAMILY_MEMBERS
COUNT()[name="COUNT"]
    MISSING=LISTWISE REPORTMISSING=NO
  /GRAPHSPEC SOURCE=INLINE.
BEGIN GPL
  SOURCE: s=userSource(id("graphdataset"))
  DATA: NUM_FAMILY_MEMBERS=col(source(s),
name("NUM_FAMILY_MEMBERS"), unit.category())
  DATA: COUNT=col(source(s), name("COUNT"))
  COORD: polar.theta(startAngle(0))
  GUIDE: axis(dim(1))
  GUIDE: legend(aesthetic(aesthetic.color.interior),
label("NUM_FAMILY_MEMBERS"))
  SCALE: linear(dim(1), dataMinimum(),
dataMaximum())
  SCALE: cat(aesthetic(aesthetic.color.interior),
sort.natural())
  ELEMENT:
interval.stack(position(summary.percent(summary.perc
ent(COUNT,
    base.all(acrossPanels())))),
color.interior(NUM_FAMILY_MEMBERS))
END GPL.
```

**TABLE 7**: Number of family members

Table 7 has shown the number of family members from 0 to 13 persons. The highest percentage of the number of family members is 6 persons with 24.9% and the lowest percentages of the number of family members are 13 persons with 0.3%. The breakdown of the relationship between number of family members and each SPM's subject is evidently shown in Table 8.

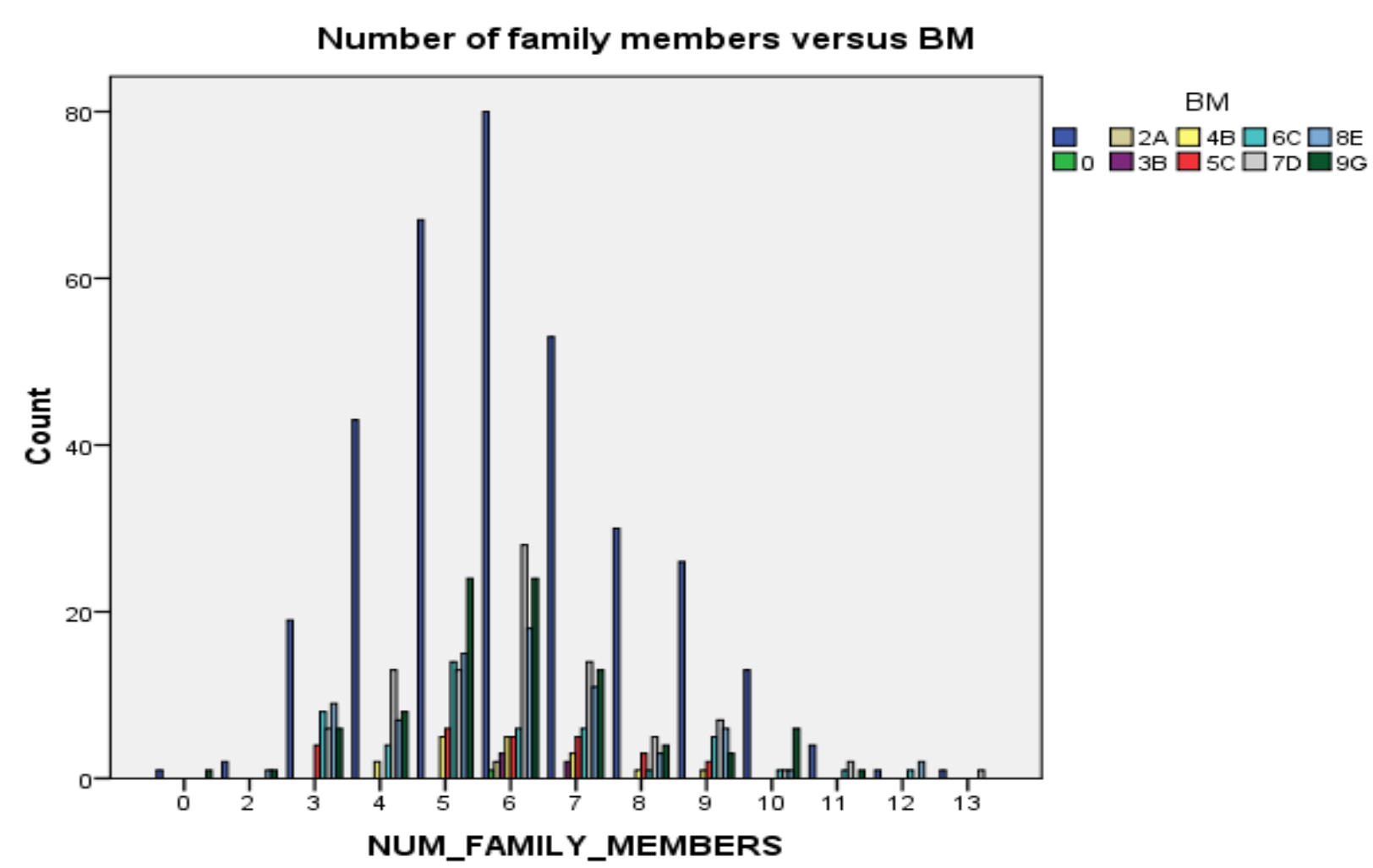

**Syntax SPSS :**

```
CROSSTABS
  /TABLES=BM_SPM BY NUM_FAMILY_MEMBERS
  /FORMAT=AVALUE TABLES
  /CELLS=COUNT
  /COUNT ROUND CELL
  /BARCHART.
```

**TABLE 8**: Number of family members versus BM

The number of family members of 6 persons shows their contribution in score with all grades for subject Bahasa Malaysia (BM). There are 2 students that get grade 2A, 3 students get grade 3B, 5 students get grade 4B, 5 students get grade 5C, 6 students get grade 6C, 28 students get grade 7D, 18 students get grade 8E and 24 students get grade 9G. The overall of this group of number family members of 6 persons was contribute for score in all grades are around 92 students. This process is automated generate using Syntax SPSS in SPSS 16.0.

This process need to continue for attributes number of family member still learning, number of family member receive SPBT, family income and SPBT. But it will not be show in this study because this study just wants to proof of focus on automated analysis that are automatically generate using Syntax SPSS and will be publish on web base data analysis result as show below (Fig 13 ~ Fig 14).

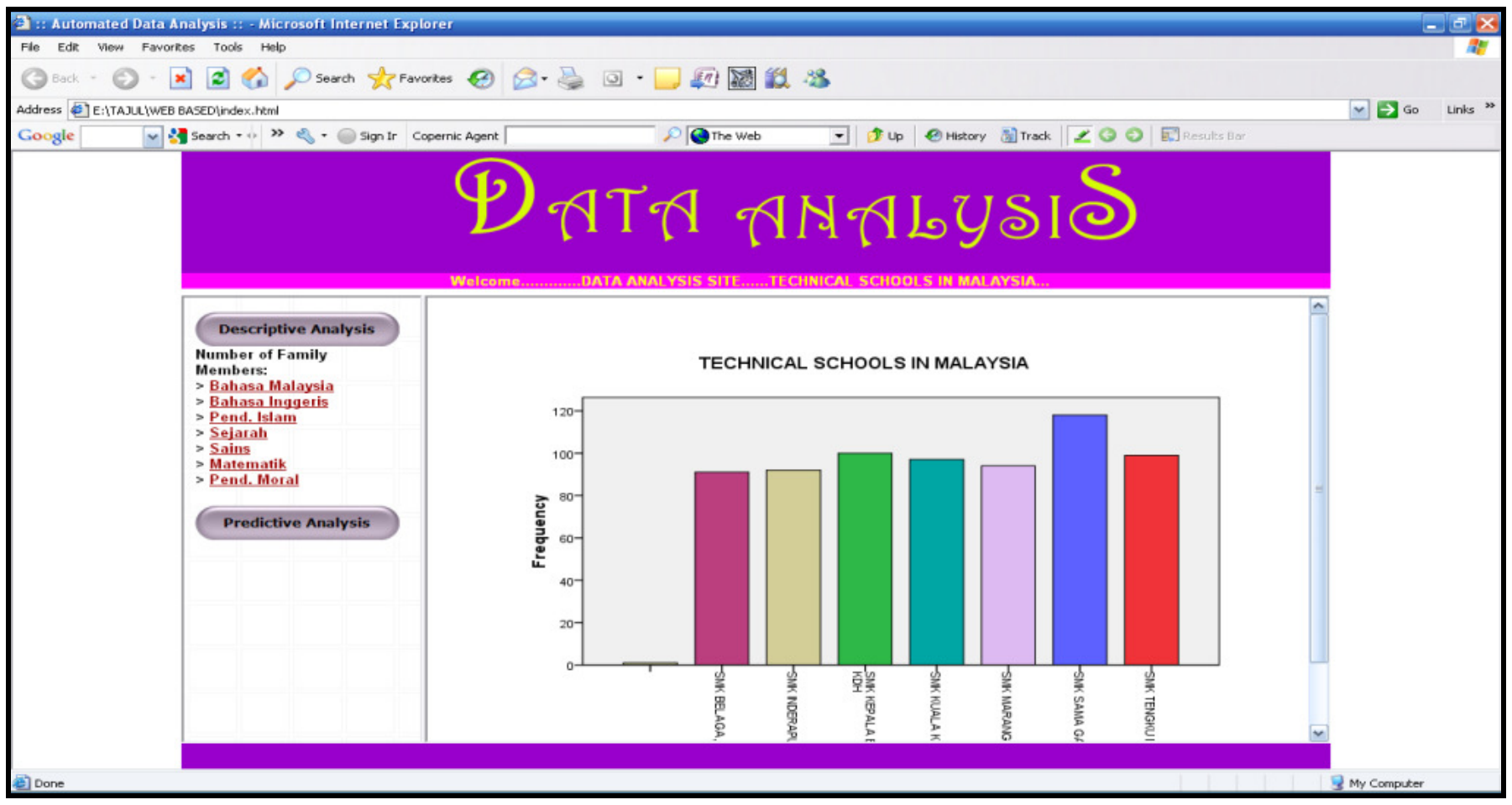

**FIGURE 13**: Main page on Data Analysis web based.

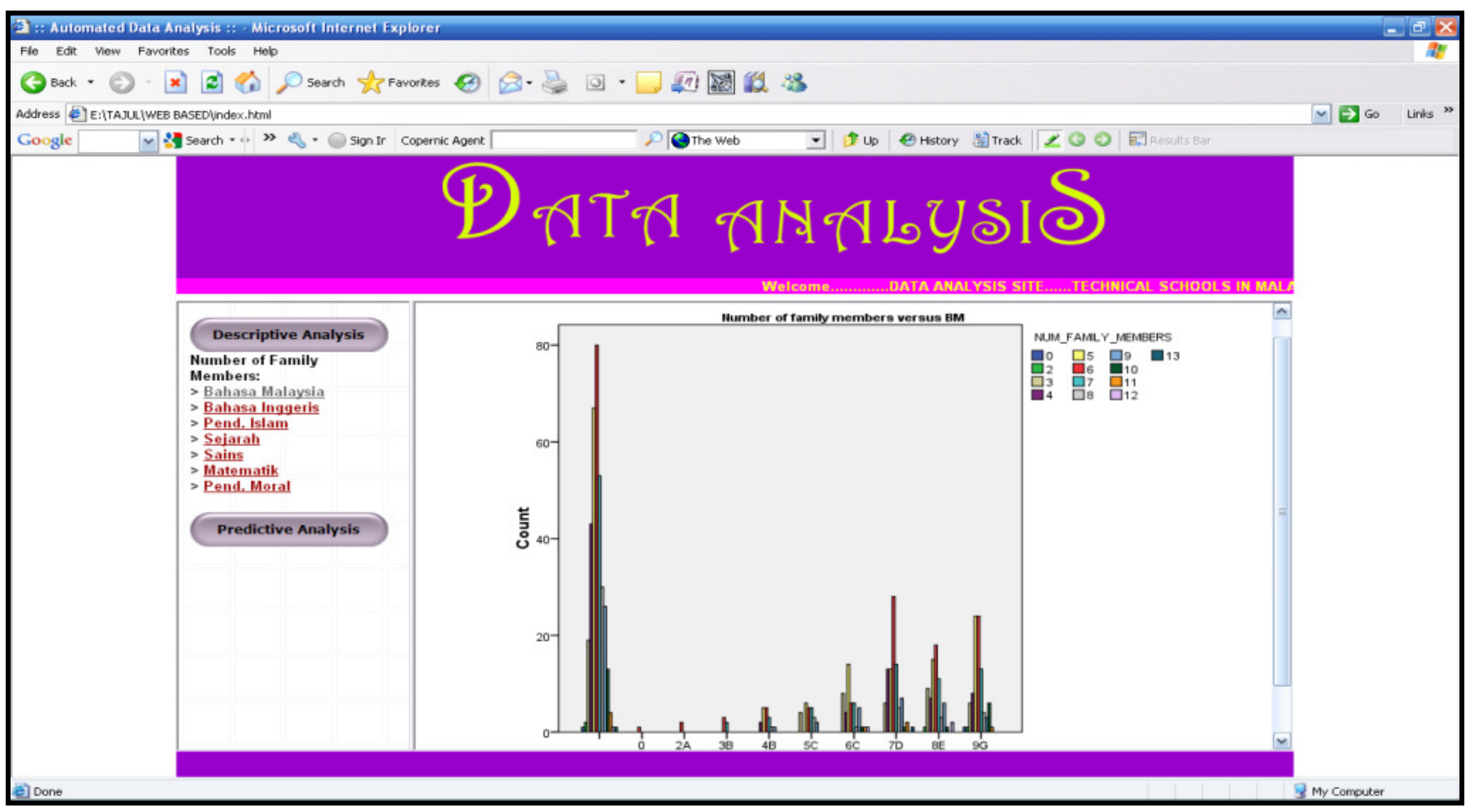

**FIGURE 14**: Description analysis on Number of family members versus BM

## 6. CONCLUSION

This study summarized the examination factors such as exam results and other factors such as SPM's subject, number of family members, number of family member still learning, number of family member receive SPBT, family income and SPBT that contribute to students' academic achievement in the future. These all factors show strong relationship between each other. The descriptive analysis was used to describe the frequency and cross tabulation between variables in this study.

As this study has illustrated, there is a potential of developing a system for centralizing the secondary technical schools student data. It also possible to perform automated descriptive analysis online that could reduce the time required to process the marks and perform manual analysis. The study also demonstrates that it is possible to integrate the proposed system with statistical analysis package in order to deliver intelligent business solutions.

Another finding from the study indicates that the analysis obtained could be used by the school management to make a suitable plan for their students' academic achievement program in the future. In addition, other data mining techniques like association rule also can be used to measure the association between attributes. The finding could be used to further enhance the strength of each attribute with description variables and among attributes or independent variables.

## 7. ACKNOWLEDGMENT

This presented study is supported by the Jabatan Pelajaran Negeri Pulau Pinang for their support direct and indirect in order to complete this research.